\documentclass[conference, a4paper]{./IEEEtran}

\ifCLASSINFOpdf
  \usepackage[pdftex]{graphicx}
\else
 
\fi

\usepackage{subfigure}
\usepackage{amsmath}
\usepackage{amssymb}
\usepackage{amsfonts}
\usepackage{multicol}

\usepackage{color}
\usepackage{flushend}

\begin{document}

\title{Performance Evaluation of Bit Division Multiplexing combined with Non-Uniform QAM}

%\author{\IEEEauthorblockN{Hugo M{\'e}ric\IEEEauthorrefmark{1}\IEEEauthorrefmark{2} and
%Jos{\'e} Miguel Piquer\IEEEauthorrefmark{2}\\
%}
%\IEEEauthorblockA{\IEEEauthorrefmark{1}INRIA Chile, Santiago, Chile}
%\IEEEauthorblockA{\IEEEauthorrefmark{2}NIC Chile Research Labs, Santiago, Chile\\
%Email: hugo.meric@inria.cl, jpiquer@nic.cl}}

\author{\IEEEauthorblockN{Hugo M{\'e}ric}
\IEEEauthorblockA{Inria Chile - NIC Chile Research Labs\\
Santiago, Chile\\
Email: hugo.meric@inria.cl}
\and
\IEEEauthorblockN{Jos{\'e} Miguel Piquer}
\IEEEauthorblockA{NIC Chile Research Labs\\
Santiago, Chile\\
Email: jpiquer@nic.cl}}

\maketitle

\begin{abstract}
Broadcasting systems have to deal with channel variability in order to offer the best spectral efficiency to the receivers. However, the transmission parameters that maximise the spectral efficiency generally leads to a large link unavailability. In this paper, we study analytically the trade-off between spectral efficiency and coverage for various channel resource allocation strategies when broadcasting two services. More precisely, we consider the following strategies: time sharing, hierarchical modulation and bit division multiplexing. Our main contribution is the combination of bit division multiplexing with non-uniform QAM to improve the performance of broadcasting systems. The results show that this scheme outperforms all the previous channel resource allocation strategies.
\end{abstract}

\IEEEpeerreviewmaketitle

\section{Introduction}

Broadcasting systems are designed to optimise the spectral efficiency and the coverage. We define the coverage as the fraction of receivers that can decode the transmitted signal. To increase the transmission data rate at a given bandwith, it is possible to increase the modulation order or the code rate. However, both solutions also decrease the coverage. Thus there is a trade-off between the spectral efficiency and the coverage.  

The first solution for broadcasting was to design the system for the worst-case reception. In that case, the coverage is 100\% but many receivers do not exploit their full potential leading to poor performance. In \cite{cover} and \cite{bergmans}, two resource channel allocations were proposed to improve the performance: time division multiplexing with variable coding and modulation, and superposition coding. Time division multiplexing, or time sharing, allocates a proportion of time to communicating with each receiver using any modulation and error protection level. Many modern broadcasting standards, for instance DVB-SH\footnote{Digital Video Broadcasting - Satellite services to Handhelds} and DVB-S2\footnote{Digital Video Broadcasting - Satellite - Second Generation}, mainly rely on time sharing \cite{sh,s2}.

\begin{figure}[!t]
\centering
\includegraphics[width = 0.74\columnwidth]{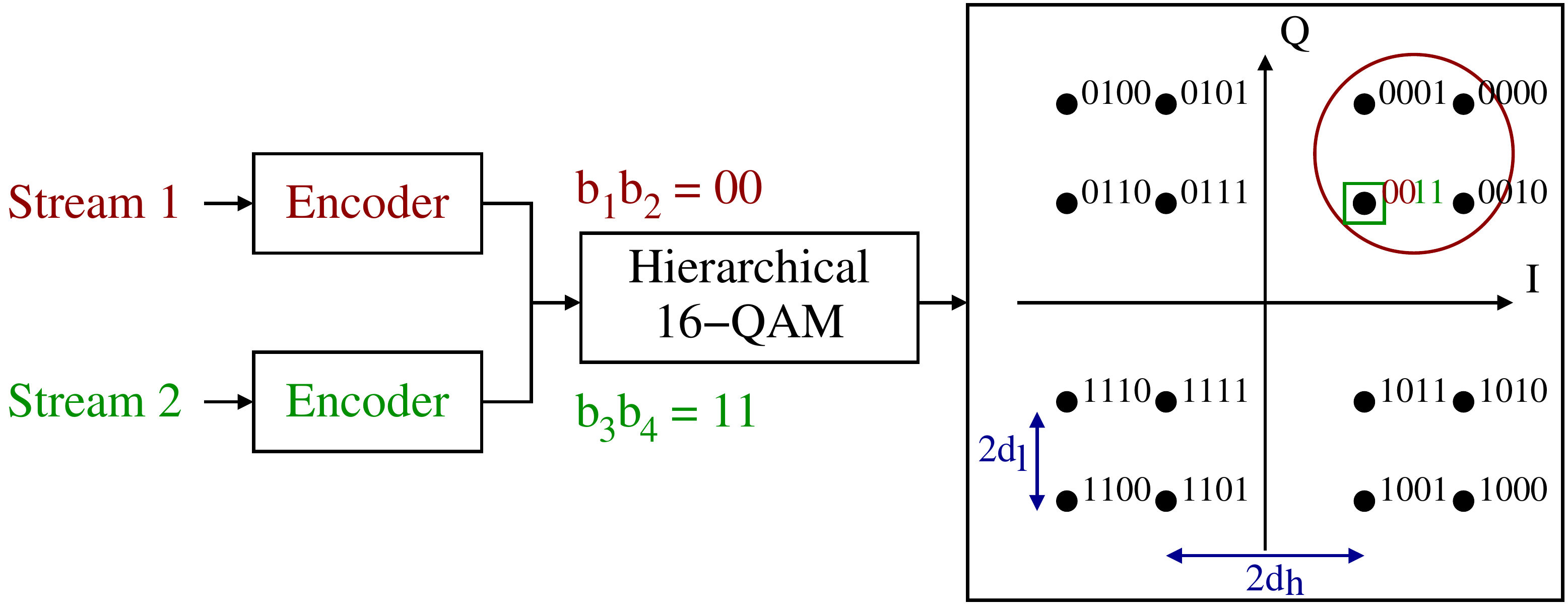}
\caption{Hierarchical modulation using a 16-QAM. Each constellation symbol carries data from two independently encoded streams.}
\label{hm_principle}
\end{figure}

In superposition coding, the available energy is shared among several service flows which are sent simultaneously in the same band. This scheme was introduced by Cover in order to increase the transmission rate from a single source to several receivers \cite{cover}. When communicating with two receivers, the principle is to superimpose information for the user with the best SNR. This superposition can be done directly at the Forward Error Correction (FEC) level or at the modulation level as shown in \figurename~\ref{hm_principle} with a 16 Quadrature Amplitude Modulation (16-QAM).

Hierarchical modulation is a practical implementation of superposition coding. The trade-off between spectral efficiency and coverage for hierarchical and non-hierarchical modulations has already been studied from theoretical and practical points of view in \cite{perf_hm} and \cite{vtc13}, respectively. The first work focuses on QAM while the second focuses on Amplitude and Phase-Shift Keying (APSK) modulations. In both cases, the results point out that hierarchical modulation may provide significant gains compared to the classical time sharing strategy. The performance depends on several parameters of the broadcasting system, e.g., the targeted coverage.

Recently, a new channel resource allocation has been introduced \cite{bdm}. This scheme, called Bit Division Multiplexing (BDM), extends the multiplexing from symbol level to bit level. In their work, the authors propose two bit allocation strategies in order to optimise the transmission rate or the decoding threshold when communicating two services. The results presented in \cite{bdm} mainly use the uniform 256-QAM.

In this paper, we study the combination of BDM with non-uniform 16-QAM for broadcasting two services. Our work focuses on the 16-QAM as it only requires one  parameter to describe the constellation geometry. Moreover, this is a preliminary study in order to evaluate the potential of BDM combined with non-uniform modulations. Analysing the trade-off between spectral efficiency and coverage, the results show that this combination outperforms all the previous schemes. Our work also completes \cite{bdm} as we present the BDM performance in terms of spectral efficiency and coverage.
 
The paper is organised as follows: Section~\ref{part2} presents the channel resource allocation strategies considered in this paper. In Section~\ref{part3}, we present and compare the previous strategies in terms of spectral efficiency and link unavailability. Finally, Section~\ref{part4} concludes the paper by summarising the results.

\section{Channel resource allocation strategies}\label{part2}

\begin{figure*}[!t]
\centering
\subfigure[Time division multiplexing]{
\includegraphics[keepaspectratio=true, width=0.315\textwidth]{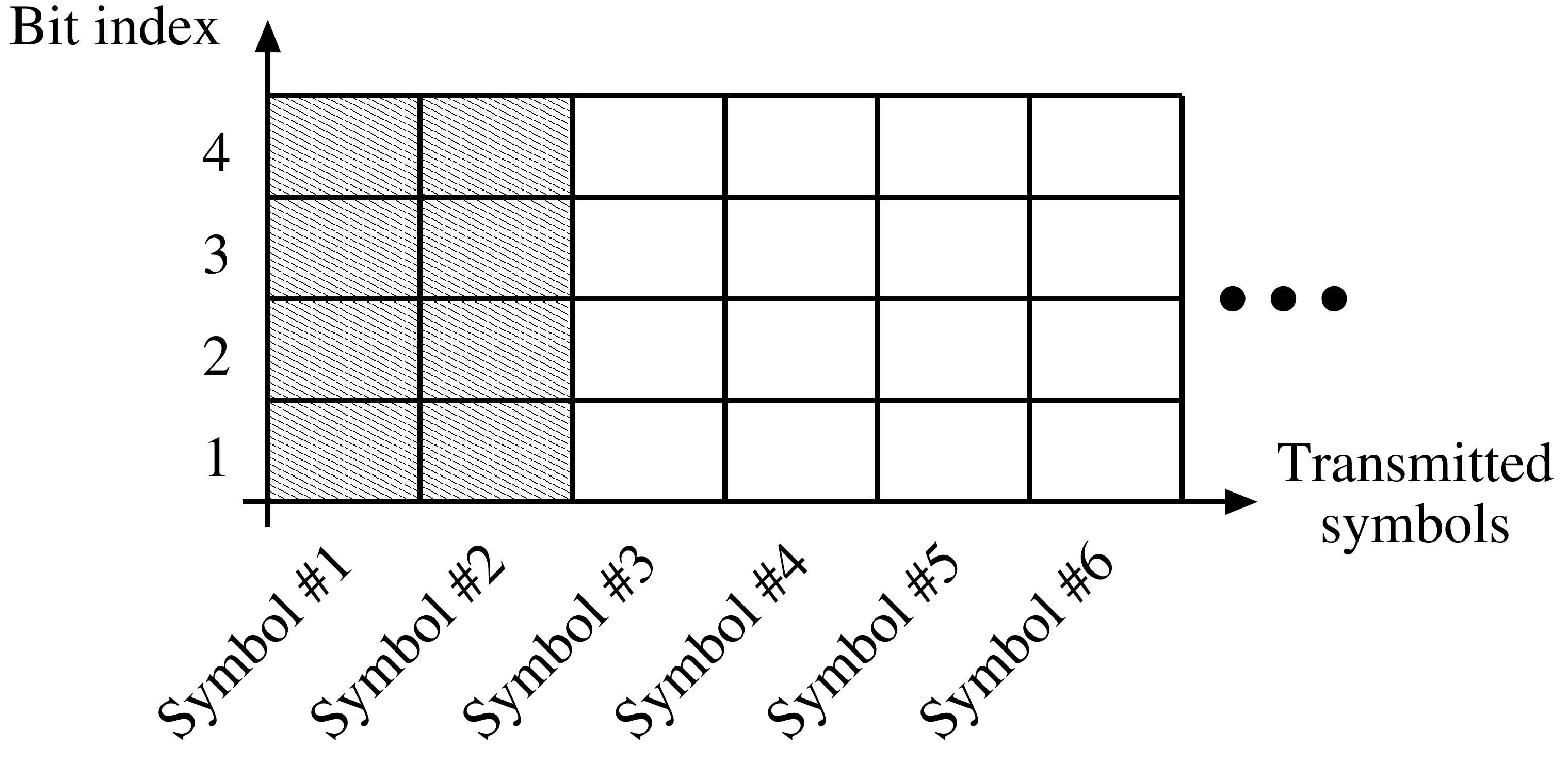}
\label{bdm_ts}
}
\subfigure[Hierarchical modulation]{
\includegraphics[keepaspectratio=true, width=0.315\textwidth]{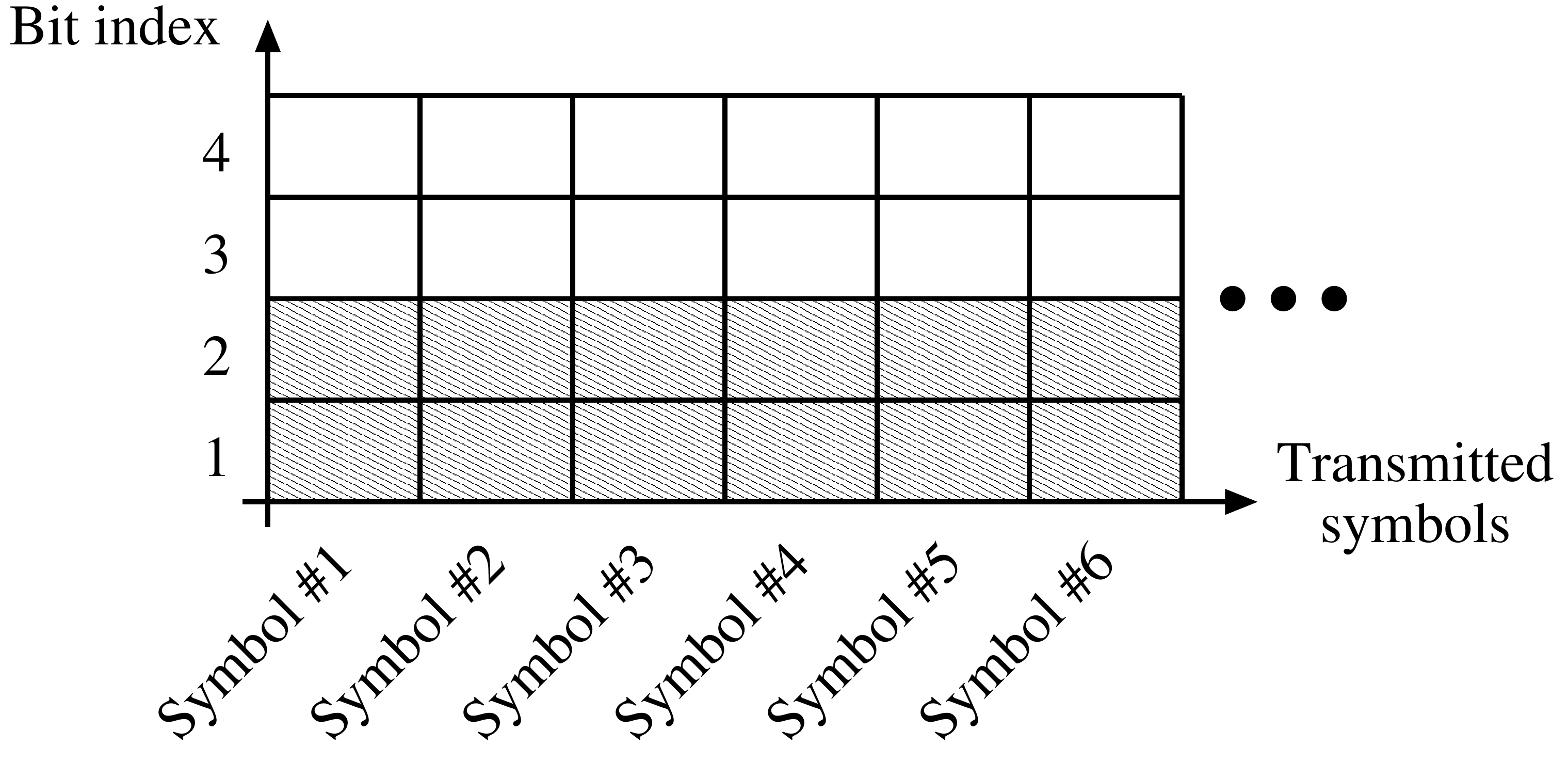}
\label{bdm_hm}
}
\subfigure[Bit division multiplexing]{
\includegraphics[keepaspectratio=true, width=0.315\textwidth]{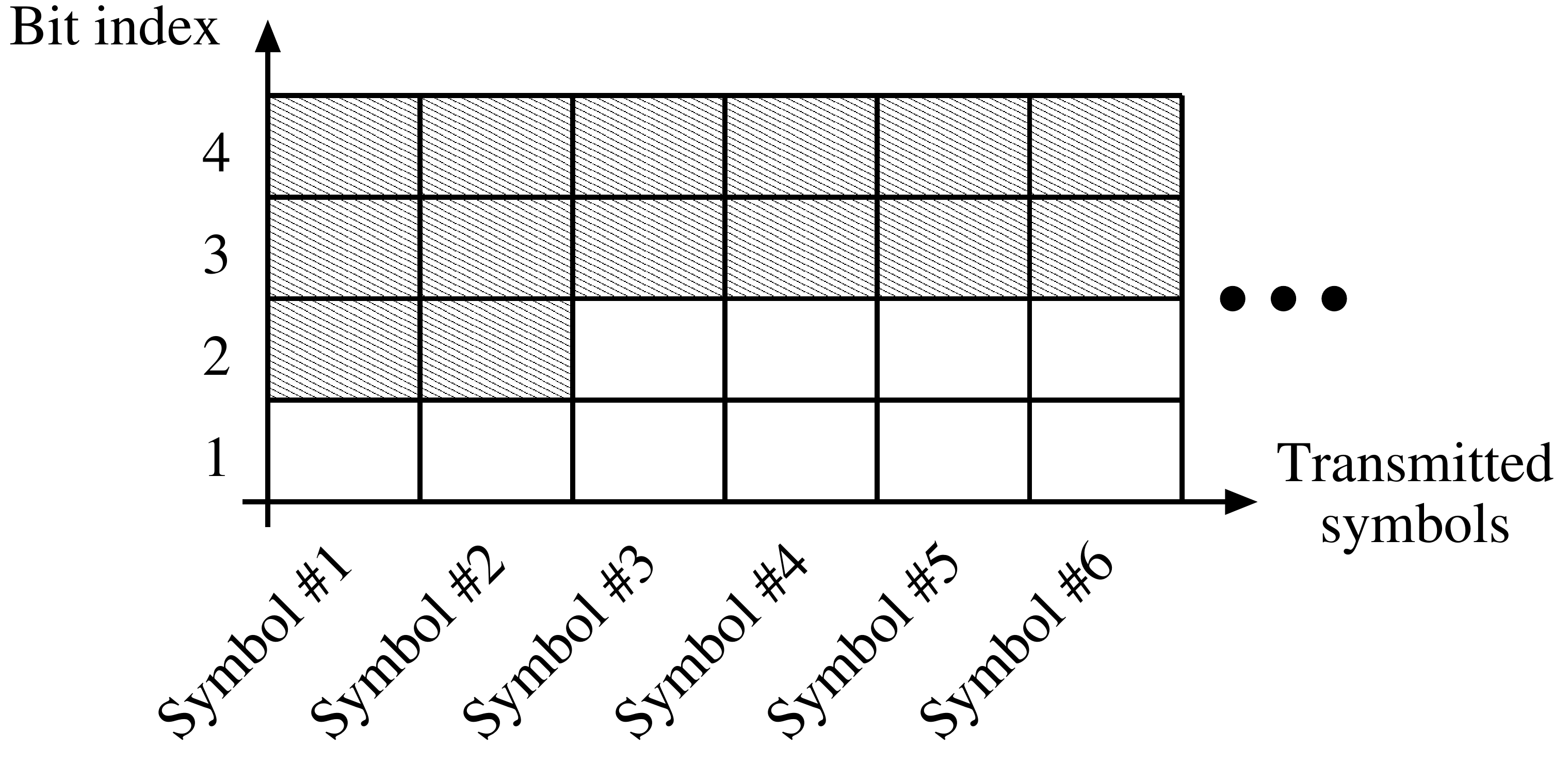}
\label{bdm_general}
}
\caption{Illustration of different channel resource allocation strategies where each transmitted symbol carries 4 bits}
\label{channel_resource_allocation}
\end{figure*}

%%%%%%%%%%%%%%%%%%%%%
%%% System definition
%%%%%%%%%%%%%%%%%%%%%
\subsection{System definition}
As already mentioned, we focus on broadcasting two services: the \emph{base stream} and the \emph{enhanced stream}. The base stream must be decoded by most receivers to ensure a relatively large coverage, while the enhanced stream is dedicated to receivers with a good channel quality. Such system can broadcast SVC encoded video \cite{svc} where the base layer of the video is transmitted in the base stream and the enhanced layer(s) in the enhanced stream.

To transmit the base and enhanced streams, several multiplexing methods can be used. We now present the three channel resource allocation strategies considered in this paper: time division multiplexing, hierarchical modulation and bit division multiplexing.

%%%%%%%%%%%%%%%%
%%% Time sharing
%%%%%%%%%%%%%%%%
\subsection{Time division multiplexing}
One of the easiest channel resource allocation is time division multiplexing or time sharing. When communicating with multiple receivers, the principle is to allocate a fraction of time for transmitting data to each receiver using any modulation and error protection level. Thus time sharing allocates resource at the symbol level. For instance, \figurename~\ref{bdm_ts} illustrates a time sharing strategy where 2 symbols out of every 6 symbols are allocated to a receiver. In that example, each symbol carries 4 bits so the modulation used for transmitting has an order of 16, for instance a 16-QAM or a 16-APSK.

Time sharing is the most use in practice today as it is very easy to implement. However, Cover proves that the theoretical transmission rates of a broadcasting system are achieved through superposition coding that may clearly outperform time division multiplexing \cite{cover}.

%%%%%%%%%%%%%%%%%%%%%%%%%%%
%%% Hierarchical modulation
%%%%%%%%%%%%%%%%%%%%%%%%%%%
\subsection{Hierarchical modulation and non-uniform constellations}
Another well-known channel resource allocation is hierarchical modulation. As mentioned before, hierarchical modulation is an implementation of superposition coding that merges several streams in a same symbol. In our study, two streams are considered. \figurename~\ref{hm_principle} depicts the hierarchical 16-QAM and the mapping used in this paper: one stream (stream 1) is transmitted with the bits $b_1$ and $b_2$, while the other stream (stream 2) is transmitted with the bits $b_3$ and $b_4$. In that case, \figurename~\ref{bdm_hm} shows how the resources are allocated.

Hierarchical modulations often rely on non-uniform constellations where the symbols are not uniformly distributed in the space. The geometry of non-uniform modulations is described using the constellation parameter(s). For the 16-QAM, the constellation parameter $\alpha$ is defined by $d_h/d_l$, where $2d_h$ is the minimum distance between two constellation points in different quadrants and $2d_l$ is the minimum distance between any constellation point (see \figurename~\ref{hm_principle}). By definition, we have $\alpha \geqslant 0$, where $\alpha=1$ corresponds to the uniform 16-QAM. At a given energy per symbol ($E_s$), when $\alpha$ grows, the constellation points in each quadrant become farther from the I and Q axes. However, the points in a same quadrant become closer. 

Generally, the bits $b_1$ and $b_2$ exhibit lower Bit Error Rate (BER) than $b_3$ and $b_4$ \cite{rec_ber}. With the previous definitions, the stream transmitted with $b_1$ and $b_2$ (stream 1 in \figurename~\ref{hm_principle}) corresponds to the base stream, while the other stream corresponds to the enhanced stream as it usually requires a better channel quality to be decoded error-free. The constellation parameter allows the transmitter to modify the decoding threshold of both streams. To illustrate a possible application, a study of scalable video broadcasted with hierarchical modulation is done in \cite{svc_hm}.

%%%%%%%%%%%%%%%%%%%%%%%%%%%%%
%%% Bit division multiplexing
%%%%%%%%%%%%%%%%%%%%%%%%%%%%%
\subsection{Bit division multiplexing}
A novel multiplexing method called bit division multiplexing has recently been introduced \cite{bdm}. BDM allocates channel resource at the bit level as depicted in \figurename~\ref{bdm_general}. By taking advantage of the unequal error protection of the bits within any constellation symbol, this strategy improves the throughput of broadcasting systems that transmit multiple services simultaneously. 

Due to its resource allocation strategy, BDM is much more flexible than time sharing and hierarchical modulation. Indeed, time division multiplexing can modify the data rate of a stream but cannot change its decoding performance, while hierarchical modulation works the opposite way. With the BDM resource allocation, the transmitter can adjust both the throughput and the decoding performance of a stream. Thus BDM extends the two previous schemes.   

In \cite{bdm}, the authors study BDM with a uniform 256-QAM in terms of transmission rates and decoding thresholds. Depending on the parameter to optimise, several resource allocation strategies have been introduced.

We propose here to combine BDM with the non-uniform 16-QAM and study the trade-off between spectral efficiency and coverage of this new scheme. The usage of non-uniform constellations allows us to vary the error protection level of the transmitted bits.

\section{Performance evaluation}\label{part3}
This section studies the trade-off between spectral efficiency and coverage for the previous channel resource allocation schemes. We begin by introducing the mutual information for QAM, then we introduce an example of receivers SNR distribution and finally we present the simulations results.

%%%%%%%%%%%%%%%%%%%%%%%%%%%%%%
%%% Mutual information for QAM
%%%%%%%%%%%%%%%%%%%%%%%%%%%%%%
\subsection{Mutual information for QAM}

For theoretical capacity-achieving codes, the mutual information is equivalent to the spectral efficiency. With the Bit Interleaved Coded Modulation assumption \cite{bicm}, each transmitted bit can be modeled as a binary-input channel. The $i$-th channel is the channel associated to the transmission of the bit in $i$-th position within the constellation mapping. The capacity of the $i$-th channel, i.e., the mutual information between the corresponding input bit and the channel output, is given by
\begin{equation}
C_i = I \left( b;y | S=i \right) = 1 - \mathbb{E}_{b,y} \left[ \log_{2} \frac{\sum_{x \in \mathcal{X}} \Pr(y|x)}{\sum_{x \in \mathcal{X}_{b}^{i}} \Pr(y|x)} \right] ,
\label{bwc}
\end{equation}
where $I \left( b;y | S=i \right)$ is the conditional mutual information; $b$, $S$ and $y$ are random variables representing the bit value, the bit index and the channel output; $\mathcal{X}$ is the set of all constellation points; $\mathcal{X}_{b}^{i} = \{ x \in \mathcal{X} | \text{the $i$-th bit of $x$ label is $b$} \}$; $\Pr$ denotes the probability and $\mathbb{E}_{b,y}$ the expectation over $b$ and $y$ \cite{perf_hm}. 

\begin{figure}[!ht]
\centering
\includegraphics[width = 0.875\columnwidth]{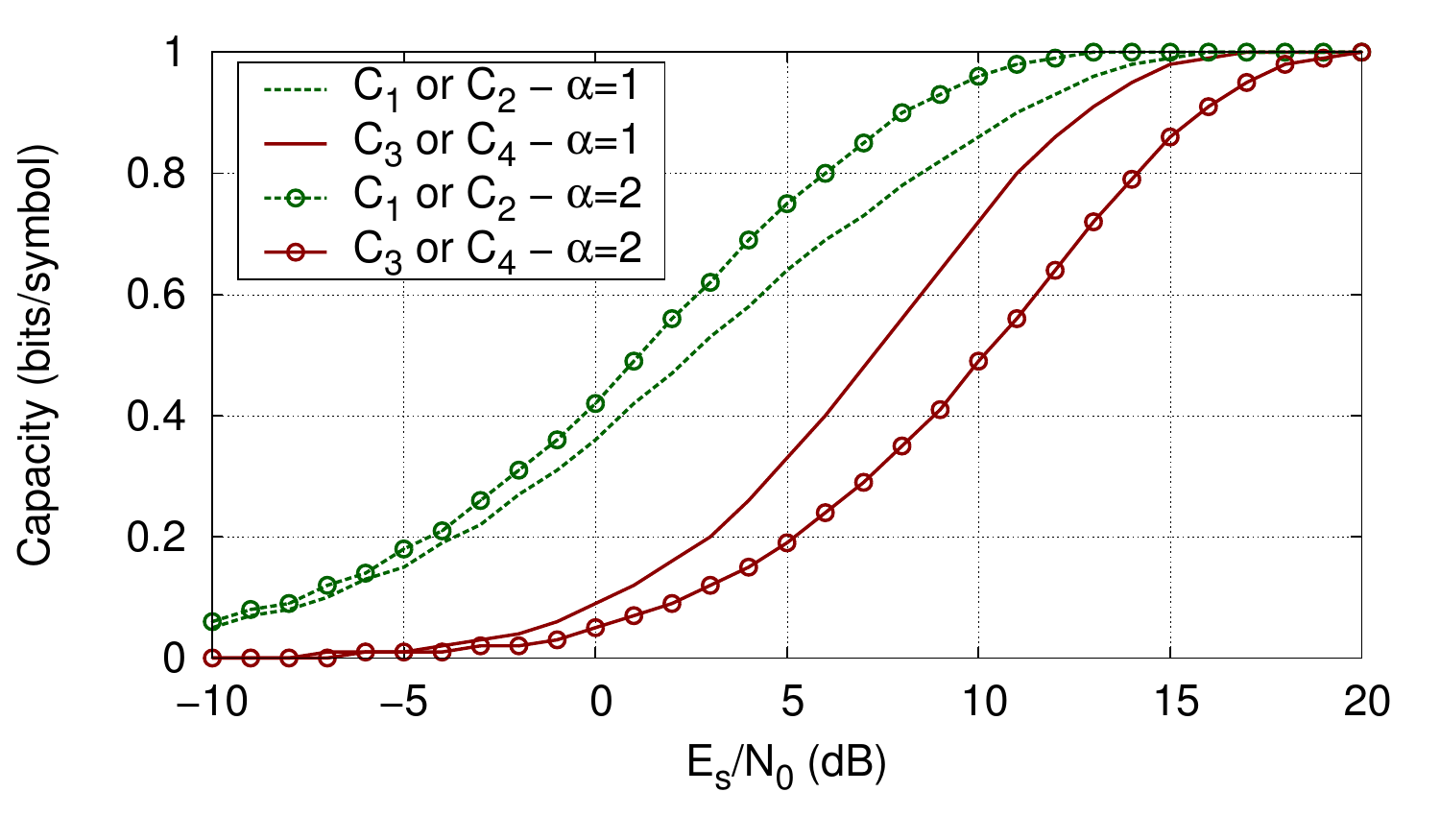}
\caption{Capacity for 16-QAM with $\alpha=1,2$: $C_i$ ($1 \leqslant i \leqslant 4$) is the capacity of the binary-input channel associated to the bit with index $i$ defined in (\ref{bwc})}
\label{capacity_qam}
\end{figure}

\figurename~\ref{capacity_qam} illustrates the capacity in (\ref{bwc}) for non-uniform 16-QAM with $\alpha$ equals to 1 and 2. At a given $E_s/N_0$, when the constellation parameters grows, we observe that $C_3$ and $C_4$ decrease while $C_1$ and $C_2$ increase. This is explained by the impact of $\alpha$ on the constellation geometry as described in Section~\ref{part2}.

Due to the resource allocation strategy, the channel is divided in several sub-channels. The capacity of a sub-channel is computed by adding the capacities of the bits composing the sub-channel \cite{bdm}. For instance in \figurename~\ref{bdm_general}, the system allocates the bits with indexes 3 and 4 plus one third of the bits with index 2 to the sub-channel marked in gray, thus the capacity of the sub-channel is
\begin{equation}
C_2/3 + C_3 + C_4,
\end{equation}
where $C_i$ is defined in (\ref{bwc}).

%%%%%%%%%%%%%%%%%%%%
%%% SNR distribution
%%%%%%%%%%%%%%%%%%%%
\subsection{Receivers SNR distribution}
The computation of the coverage requires to know the receivers SNR distribution. We use here the model in \cite{perf_hm} based on IEEE 802.16m. Considering a circular cell, a receiver at a distance $r$  from the transmitter has a SNR given by
\begin{equation}
\text{SNR}_{\text{dB}} = 10.81 - 37.6 \log_{10}(r) - X_\sigma ,
\end{equation}
where $r$ is in kilometers, $X_\sigma$ is a Gaussian random variable with zero mean and standard deviation $\sigma=8$ dB.

We assume that the receivers are uniformly distributed in the cell and the cell radius is 0.75 km. Thus it is possible to compute the SNR to obtain a targeted coverage. We focus on several coverage values that are resumed in Table~\ref{snr_vs_coverage} with their associated SNR thresholds. For a given coverage, the spectral efficiency is defined as the mutual information at the SNR associated to the coverage.
\begin{table}[!ht]
\renewcommand{\arraystretch}{1.1}
\caption{SNR thresholds associated to different coverage values}
\label{snr_vs_coverage}
\centering
\begin{tabular}{c||c|c|c|c|c} 
\hline
Coverage (\%) & 98 & 95 & 90 & 80 & 70 \\
\hline
SNR (dB) & 3.4 & 7 & 10.3 & 14.4 & 17.4 \\
\hline 
\end{tabular}
\end{table}

%%%%%%%%%%%%%%%%%%%%%%%%%%
%%% Performance evaluation
%%%%%%%%%%%%%%%%%%%%%%%%%%
\subsection{Performance evaluation}

%%% Hypotheses
\textbf{Hypotheses.} We consider a system that broadcasts a base stream and an enhanced stream. The targeted coverage for each stream is set in the simulations.

The system implements one of the following resource allocation strategies: time division multiplexing, hierarchical modulation, BDM with or without non-uniform constellations.

Finally, the transmitted signal is modulated with a 16-QAM. The time sharing and BDM strategies only use uniform 16-QAM, i.e., 16-QAM with $\alpha=1$.

%%% Bit allocation for BDM strategies
\textbf{Bit allocation for BDM strategies.} For the strategies based on BDM, many bit allocations are possible. In our study, we set the coverage of each stream and search to maximise the spectral efficiency. As the receiver SNR distribution is known, setting the coverage is equivalent to set the decoding thresholds for both streams. It is important to remark that the larger the coverage, the lower the decoding threshold.

With the previous remarks, Proposition~1 in \cite{bdm} explains how to allocate the bits to the base and enhanced streams. We note $C_i^b$ and $C_i^e$ the capacities in (\ref{bwc}) evaluated at the decoding thresholds of the base and enhanced streams, respectively. Then the bit allocation is as follows: the bits with lower value of $C_i^e/C_i^b$ are allocated to the base stream and the rest bits to the enhanced stream.

%%% Spectral efficiency computation

\begin{figure*}[!t]
\centering
\subfigure[$G_b=98\% , G_e=90\%$]{
\includegraphics[keepaspectratio=true, width=0.45\textwidth]{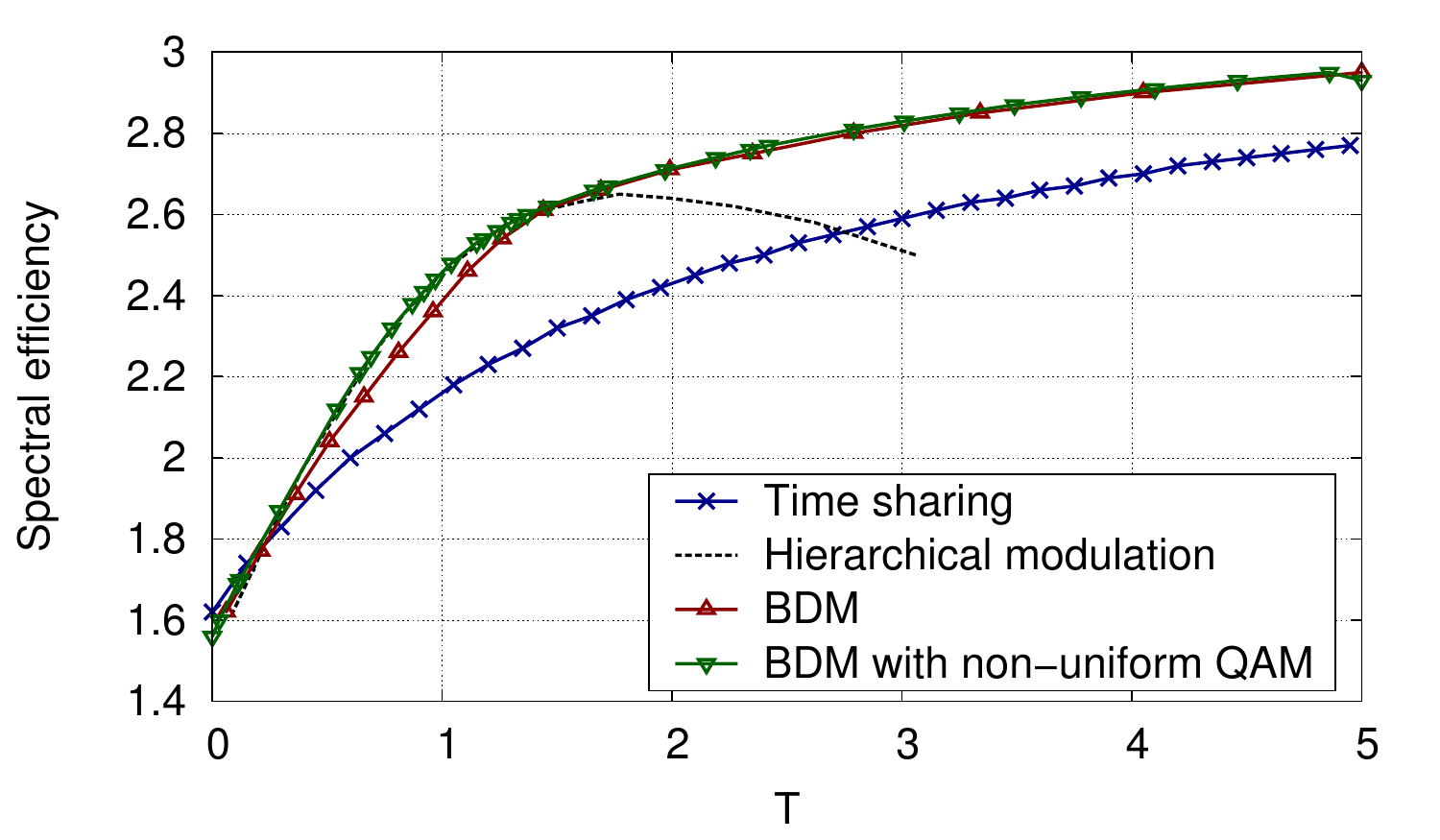}
\label{result1}
}
\subfigure[$G_b=98\% , G_e=80\%$]{
\includegraphics[keepaspectratio=true, width=0.45\textwidth]{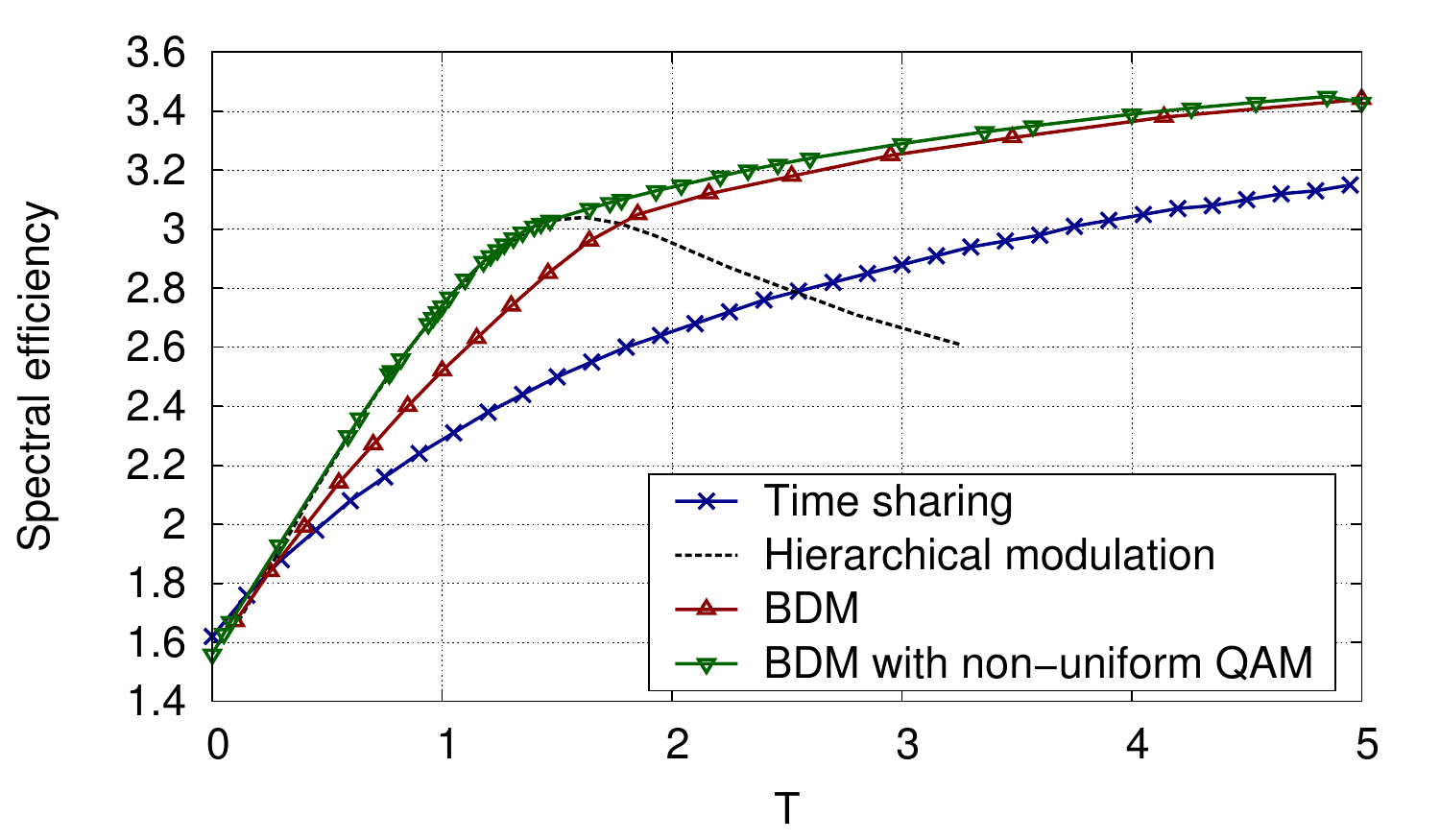}
\label{result2}
}
\subfigure[$G_b=95\% , G_e=80\%$]{
\includegraphics[keepaspectratio=true, width=0.45\textwidth]{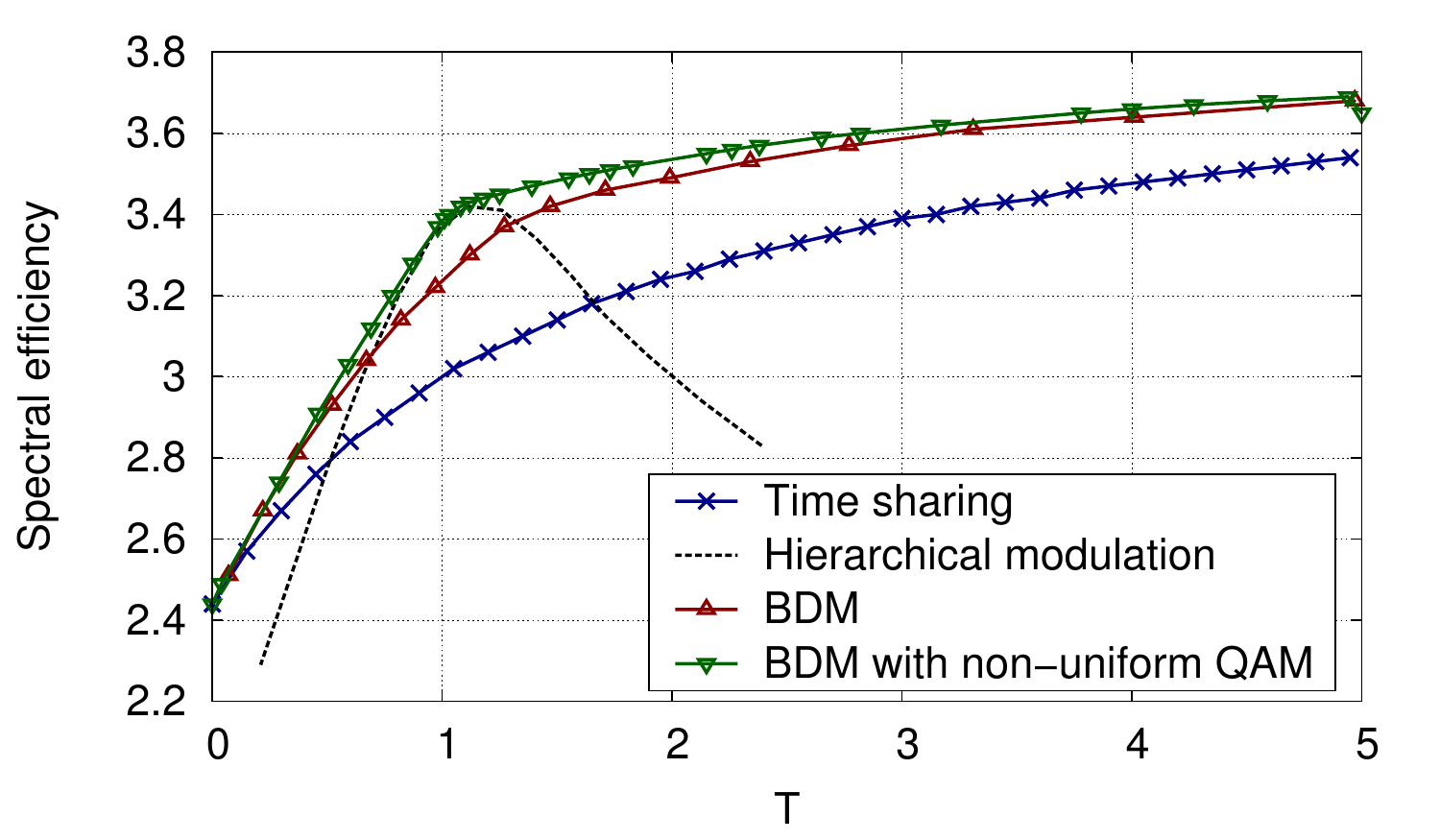}
\label{result3}
}
\subfigure[$G_b=95\% , G_e=70\%$]{
\includegraphics[keepaspectratio=true, width=0.45\textwidth]{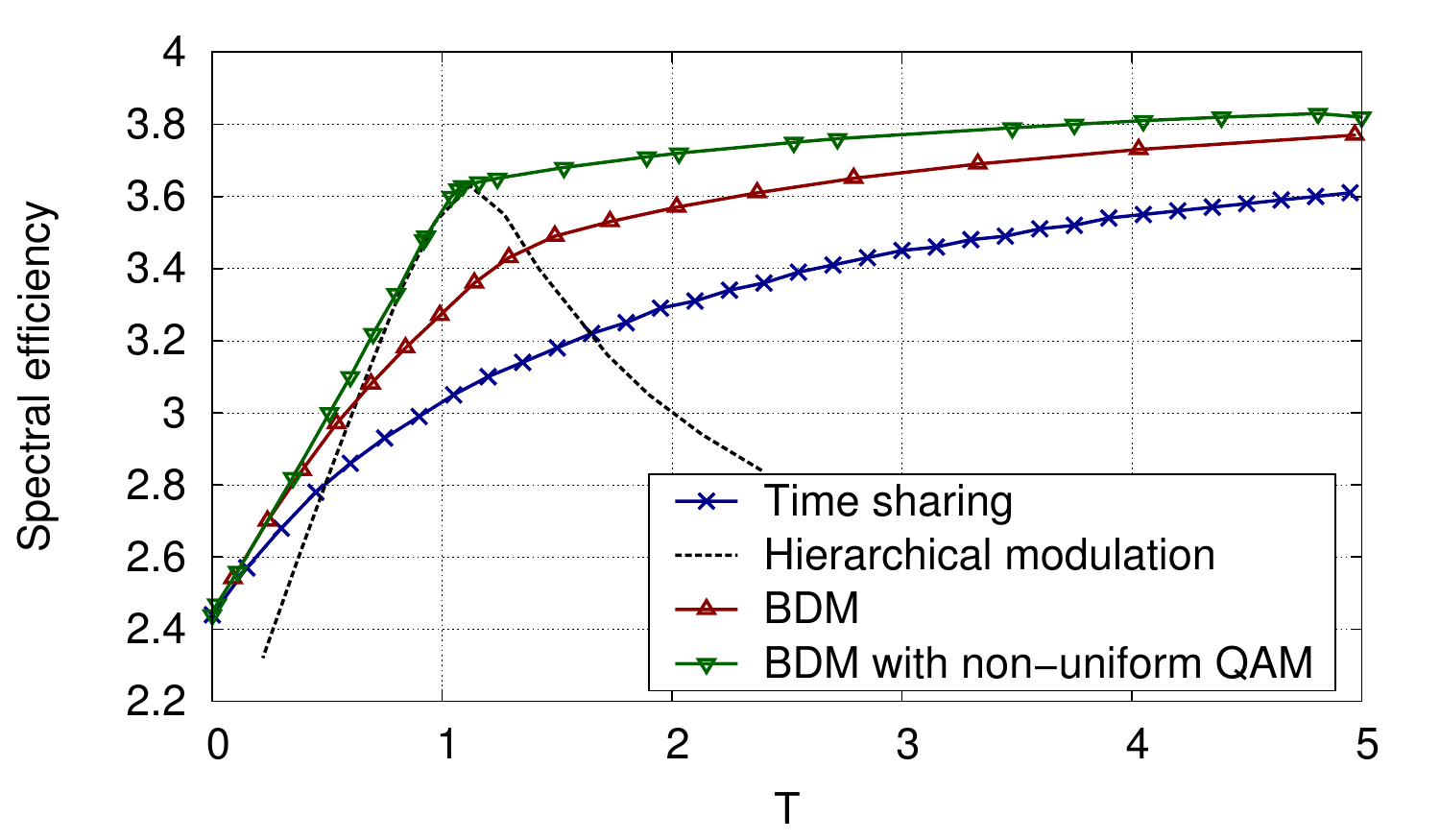}
\label{result4}
}
\caption{Spectral efficiency in function of $T$, $G_b$ and $G_e$ for the 4 channel resource allocation strategies considered in this paper}
\label{final_results}
\end{figure*}

\textbf{Spectral efficiency computation.} We evaluate the performance in terms of \emph{overall spectral efficiency} with a given coverage for both services. The overall spectral efficiency refers here to the sum of the spectral efficiencies of the base and enhanced streams.

We note $G_b$ and $G_e$ the coverage for the base stream and enhanced stream, respectively. Once the coverage $G_i$ of a stream is set, its spectral efficiency $C_i$ is given by the mutual information at the SNR associated to $G_i$ (see Table~\ref{snr_vs_coverage}).

We introduce another system parameter, denoted $T$, that is equals to the ratio between the data rate of the enhanced and base streams, i.e., $T=C_e/C_b$. Time sharing and BDM support any $T \geqslant 0$, while hierarchical modulation cannot offer $T$ larger than a specific value \cite{perf_hm}. In the rest of the paper, we are interested to obtain the (overall) spectral efficiency in function of $T$ for a given $(G_b, G_e)$ pair.

We now give an expression of the spectral efficiency for time sharing. We assume that $T$, $G_b$ and $G_e$ are set. Using the coverage values and the SNR distribution, the spectral efficiencies of the 16-QAM for the base and enhanced streams are $C_b$ and $C_e$, respectively. By definition, we have
\begin{equation}
T = \frac{x C_e}{(1-x)C_b},
\label{eqT}
\end{equation}
where $x$ is the time transmitting the enhanced stream. Solving (\ref{eqT}), we obtain
\begin{equation}
x = \frac{T C_b}{C_e + T C_b}.
\end{equation}
Finally, the spectral efficiency for time sharing is given by
\begin{equation}
x C_e + (1-x) C_b = \frac{1 + T}{1/C_b + T C_b/ C_e}.
\end{equation}

For hierarchical modulation and BDM, it is not possible to give an analytical form of the spectral efficiency as the problem involves more parameters: the bit allocation strategy and $\alpha$. Thus we rely on simulations.

%%% Simulations results
\textbf{Simulations results.} \figurename~\ref{final_results} presents the results in terms of spectral efficiency for several $(G_b, G_e)$ pairs and the 4 channel resource allocation strategies considered in this paper. For the strategies that rely on the non-uniform 16-QAM, the parameter $\alpha$ varies between 0 and 15 with a step of 0.02. Moreover, we only present the results for $T \leqslant 5$.

First, the worst results are always obtained by time sharing or hierarchical modulation. This can be explained as both strategies are a particular case of BDM with or without non-uniform constellation. Depending on the system parameters, we remark that hierarchical modulation may clearly outperforms time division multiplexing as already observed in \cite{perf_hm} and \cite{vtc13}, but also BDM without non-uniform QAM (for instance when $T \leqslant 1.5$ in \figurename~\ref{result2}).

Then, the hierarchical modulation curves are consistent with the work presented in \cite{perf_hm}. We retrieve that hierarchical modulation cannot reach large $T$ values, for instance $T_{\text{max}} \approx 3.1$ in \figurename~\ref{result1}. The upper bound of $T$ is obtained for $\alpha=0$, while increasing $\alpha$ makes $T$ decreases. Also, we verify that
\begin{equation}
T \xrightarrow[\alpha \rightarrow \infty]{} 0.
\end{equation}

As already mentioned, BDM obtains better performance that time sharing for $T \leqslant 5$. This point completes the results presented in \cite{bdm} where it was shown that BDM outperforms time sharing in terms of transmission rate. However, spectral efficiency is not the only criterion to optimise for practical systems. Our study shows and quantifies the performance improvement of BDM over time division multiplexing when considering spectral efficiency and coverage together.

As BDM combined with non-uniform constellations extends hierarchical modulation and BDM, it naturally obtains the best performance. However, \figurename~\ref{final_results} shows that the performance greatly depends on the values of $T$, $G_b$ and $G_e$. For instance, BDM with or without non-uniform constellations perform similarly when $G_b=98$ \% (see \figurename~\ref{result1} and \figurename~\ref{result2}), but gains up to 10\% are observed in \figurename~\ref{result4} for $T=1$. For large $T$ value, the difference between BDM with and without non-uniform 16-QAM vanishes. Indeed, in that case the constellation parameter $\alpha$ does not affect anymore the performance.

\section{Conclusion and future work}\label{part4}

In this paper, we combine BDM with non-uniform 16-QAM. We study the trade-off between spectral efficiency and coverage for this new scheme. The results show that the proposed scheme outperforms time sharing, hierarchical modulation and BDM. Our study completes \cite{perf_hm} as we present the results for two new channel ressource allocation strategies. It also completes \cite{bdm} as it gives the performance of BDM in terms of spectral efficiency and coverage.

Future work will investigate the performance improvement that BDM can provide in practical broadcasting systems such as DVB-SH and DVB-S2. We also plan to extend our work to higher order non-uniform modulations such as the 256-QAM.

\ifCLASSOPTIONcaptionsoff
  \newpage
\fi

\nocite{*}
\bibliographystyle{IEEEtran}
\bibliography{biblio}

\end{document}